\documentstyle[preprint,aps,floats,epsf]{revtex}
\begin{document}
\baselineskip=18pt
\def\be{\begin{equation}}
\def\ee{\end{equation}}
\def\bea{\begin{eqnarray}}
\def\eea{\end{eqnarray}}
\def\E{{\rm e}}
\def\bearst{\begin{eqnarray*}}
\def\eearst{\end{eqnarray*}}
\def\peleven{\parbox{11cm}}
\def\peffec{\peight{\bearst\eearst}\hfill\peleven}
\def\pspace{\peight{\bearst\eearst}\hfill}
\def\ptwelve{\parbox{12cm}}
\def\peight{\parbox{8mm}}

\title
{Strong Coupling Limit of the Kardar-Parisi-Zhang Equation in
$2+1$ Dimensions }
\author
{  F. Shahbazi $^{a}$,   A. A. Masoudi $^{a}$, and M. Reza Rahimi
Tabar $^{b,c}$}
\address
{\it $^a$ Department of Physics, Sharif University of Technology,
P.O.Box 11365-9161, Tehran, Iran.\\
$^b$ CNRS UMR 6529, Observatoire de la C$\hat o$te d'Azur,
BP 4229, 06304 Nice Cedex 4, France,\\
$^c$ Department of Physics, Iran  University of Science and
Technology, Narmak, Tehran 16844, Iran.\\
}


\maketitle


\begin{abstract}

A master equation for the Kardar-Parisi-Zhang (KPZ) equation in 2+1
dimensions is developed. In the fully nonlinear regime we derive
the finite time scale of the singularity formation in terms of the
characteristics of forcing. The exact probability density function
of the one point height field is obtained correspondingly.

\end{abstract}
\newpage
\section{Introduction}
 Because of technical importance and fundamental interest, a
great deal of efforts have been devoted to the understanding of
the mechanism of thin-film growth and the kinetic roughening of
growing surfaces in various growth techniques. Analytical and
numerical treatments of simple growth models suggest that, quite
generally, the height fluctuations have a self-similar character
and their average correlations exhibit a dynamical scaling form.
Numerous theoretical models have been proposed, of which the
simplest nontrivial example is the Kardar-Parisi-Zhang equation
[1]
 \be
 \frac{\partial h}{\partial t} -
\frac{\alpha}{2}(\nabla h)^2 = \nu \nabla^2 h + f(x,y,t),
 \ee

where $h(\bf x, t)$ specifies the height of the surface at point
$\bf x$ and $f$ is a zero-mean, statistically homogeneous, white
in time Gaussian process with covariance $ \langle f({\bf {x}},t)
f({\bf x'} ,t') \rangle=2 D_0 \delta(t-t') D({\bf{x-x'}})$.
Typically the spatial correlation of forcing is considered to be
as a delta function, mimicking the short range correlation. Here
we consider the spatial correlation as $ D(\bf{x-x'}) =
\frac{1}{\pi \sigma^2} \exp(- \frac{ (x-x')^2}{\sigma^2}),$
 where $\sigma$ is the variance of $D({\bf x-x'})$. When the variance
$\sigma$ is much less than the system size $L$ , i.e. $\sigma\ll
L$, the model represents a short range character for the forcing.
This same equation is believed to describe the statistics of
directed polymers in a random medium \cite{8}, where $h$ has the
meaning of the free energy.
 The average force on the interface is not
essential and can be removed by a simple shift in $h$. Every term
in the eq.(1) involves a specific physical phenomenon contributing
to the surface evolution. The parameters $\nu$, $\alpha$  and
$D_0$ ( and $\sigma$) are related to surface relaxation, lateral
growth and the noise strength, respectively. The
 effective coupling constant for the KPZ equation is given by
$ g = \frac{2 \alpha^2 D_0}{\nu^3}$. The limit $ g \rightarrow
\infty$ is known as the strong coupling limit. Despite the intense
effort in recent years, the properties of the strong coupling
phase are rather poorly understood.
 Only the critical
exponents of the strong-coupling regime ($g \rightarrow \infty $
or $\nu \rightarrow 0$) are known in 1+1 dimensions and their
values in higher dimensions as well as properties of the
roughening transition have been known only numerically
 and by the various approximative schemes.
 \\
Theoretical richness of the KPZ model is partly due to its close
relationships with other areas of statistical physics. It is shown
that there is a mapping between the equilibrium statistical
mechanics of a two dimensional smectic-A liquid crystal onto the
non-equilibrium dynamics of the (1+1)- dimensional stochastic KPZ
equation \cite{3}. It has been shown \cite{4} that, one can map
the kinetics of the annihilation process $A + B \rightarrow 0$
with driven diffusion onto the (1+1)-dimensional KPZ equation.
Also the KPZ equation is closely related to the dynamics of a
Sine-Gordon chain \cite{5}, the driven-diffusion equations
\cite{6,7}, and directed paths in the random media \cite{8} and
charge density waves \cite{9}, dislocations in disordered solids
\cite{10}, formation of large-scale structure in the universe
\cite{11,12}, Burgers turbulence \cite{13} and etc.

 In this paper we attempt to clarify the short time physics of KPZ
equation
 in the strong coupling limit by presenting an
 exact solution of the model in 2+1 dimensions. It turns out that eq.(1)
 exhibits a finite time singularity at time scale $t_c$.
  Using the master equation method
 it is
 shown that $t_c$ can be expressed in terms of the characteristics of forcing as,
  $t_c = \frac{1}{4}  ( \frac{ \pi }{8 \alpha ^2 D_0})^{1/3} \sigma ^2$.
Also we derive the time dependence of the moments $ <( h - \bar
h)^n> $ for time scales before creation of singularities.

The main properties of the KPZ equation in the strong-coupling
limit are as follow: (i) In the limit of $ \nu \rightarrow 0$ the
unforced KPZ (Burgers) equation develops singularities for the
given dimension. In one spatial dimension it develops sharp
valleys
 (point shocks). In the singular points the height gradients are not
continuous \cite{14}.
 In two spatial dimensions KPZ equation develops
three types of singularities, the first singularities are finite
sharp valley lines (shock lines) across which the height
gradients are discontinuous. The second type is the end point of
the sharp valley lines which separates the regular points and
singular region and is called
 a kurtoparabolic point. As time increases these sharp valley lines hit
each other and crossing point of two valley lines (shock lines)
produces a valley  node (shock node). Generically kurtoparabolic
points disappear at large times and only a network of sharp valley
lines survive \cite{11}.  A complete classification of the
singularities of KPZ (Burgers) equation in two and three
dimensions, by considering the metamorphoses of singularities as
time elapses, has been done in \cite{15}.
 (ii) For white in time
and smooth in space forcing, the sharp valley lines (cusp lines )
are smooth curves and the singularity structures of the forced
case is similar to unforced problem. It is shown in \cite{16}
that in the stationary state the  sharp valley lines produce a
curvilinear hexagon lattice in two dimensions and, therefore, one
finds a hexagonal tiling of singularity lines (valley lines).

Here we shall adapt the master equation approach which enables us
to investigate the KPZ equation in the strong coupling limit [14].
It is one of the advantages of this method that allows us to write
all the nonlinearities due to the nonlinear term
$\frac{\alpha}{2}(\nabla h)^2$ in a closed form. When $\sigma$ is
finite, the nonlinear term develops {\it finite time}
singularities for the height derivatives in the limit ($\nu
\rightarrow 0$). Therefore, one would distinguish between
different time regimes before and after the sharp valley
formation. Starting from a flat initial condition, i.e. $h(x,0)=0,
h_x = u(x,0)=0$ and $ h_y = v(x,0)=0$, we aim at tracing the time
evolution given by the KPZ equation with $\nu \rightarrow 0$. Only
for time scales before the sharp valley formation ($t < t_c$) the
limit $\nu \to 0$ and $\nu=0$ are equivalent. For $t > t_c$, the
anomalous contributions of the diffusion terms to the equation
of  probability distribution function
 will
 break this equivalence [14]. Therefore, for $t< t_c$, one can impose $\nu=0$ in the
dynamics and try to find realisable statistical solutions for the
problem. We show that it is possible to find finite time
realisable solutions for the probability density function of the
height field at least in the early steps of the evolution of the
surface. We attribute the finite life time of the realisable
ensemble of statistical solutions to the fact that after a finite
time the derivatives of the field $h(x,t)$ become singular and
, therefore, the diffusion term contributions are no more negligible.

\section{calculation of the height  moments before the singularity
formation}

 Let us consider  the Kardar-Parisi-Zhang equation in
2+1 dimensions,

\be\label{kpz}
 h_t(x,y,t)-\frac{\alpha}{2}(h_{x}^2+h_{y}^{2}) = \nu
(h_{xx}+h_{yy})+f(x,y,t). \ee

Defining $h_{x}(x,y,t)=u(x,y,t)$ and $h_{y}(x,y,t)=v(x,y,t)$,
differentiating the above equation with respect to $x$ and $y$  and
neglecting the viscosity term in the limit of $\nu\rightarrow 0$
before the creation of singularities we get

\bea
u_{t}(x,y,t)&=&\alpha(u(x,y,t)u_{x}(x,y,t)+v(x,y,t)u_{y}(x,y,t))+f_{x}(x,y,t)\label{b1}\\
v_{t}(x,y,t)&=&\alpha(u(x,y,t)v_{x}(x,y,t)+v(x,y,t)v_{y}(x,y,t))+f_{y}(x,y,t).\label{b2}
\eea

Now introducing  $\Theta$ as

 \bea && \Theta =\nonumber \\ &&
\exp{\left(-i\lambda(h(x,y,t)-\bar{h}(t))-i\mu_{1}u(x,y,t)-i\mu_{2}v(x,y,t)
\right)}.
 \eea
enables us to express the generating function as
$Z(\lambda,\mu_{1},\mu_{2},x,y,t)=\langle\Theta\rangle$. Using the
KPZ equation and its differentiations with respect to $x$ and $y$
 and neglecting the viscosity term, in the
limit of $\nu\rightarrow 0$ before the creation of singularities,
we get the following expression for the time evolution of $Z(\lambda,\mu_{1},\mu_{2},x,y,t)$

\bea\label{evolut}
 Z_{t}&=&i\gamma(t)\lambda
Z-i\lambda\frac{\alpha}{2} \langle u^2\Theta\rangle -
i\lambda\frac{\alpha}{2}\langle
v^2\Theta\rangle\nonumber\\
&-&i\alpha\mu_{1}\langle uu_{x}\Theta\rangle-i\alpha
\mu_{1}\langle vv_{x}\Theta\rangle -i\alpha\mu_{2}\langle
uu_{y}\Theta\rangle-i\alpha\mu_{2}\langle
vv_{y}\Theta\rangle\nonumber\\
&-&i\lambda\langle f(x,y,t)\Theta \rangle -i\mu_{1}\langle
f_{x}(x,y,t)\Theta \rangle -i\mu_{2}\langle f_{y}(x,y,t)\Theta
\rangle \eea

\noindent where $\gamma(t)=\bar h_{t}=\frac{\alpha}{2}<u^2+v^2>, k({\bf{ x-x'}})=2D_0D({\bf {x-x'}}),k''(0,0)=k_{xx}(0,0)=k_{yy}(0,0)$.
To simplify the above expression, by taking the derivative of $Z$ with respect
to  $x$ and $y$, and using  the homogeneity conditions, we obtain

\bea
Z_{x}&=&\langle(-i\lambda u-i\mu_{1} u_{x}-i\mu_{2}
u_{y})\Theta\rangle=0\label{h1}\\
Z_{y}&=&\langle(-i\lambda v-i\mu_{1} v_{x}-i\mu_{2}
v_{y})\Theta\rangle=0, \eea

from which it can be easily derived

\be\label{h2}
 i\frac{\partial}{\partial\mu_{1}}\langle(-i\mu_{1}
u_{x}-i\mu_{2} u_{y})\Theta\rangle=\langle
u_{x}\Theta\rangle-i\mu_{1}\langle
uu_{x}\Theta\rangle-i\mu_{2}\langle uu_{y}\Theta\rangle. \ee

From the eqs. ({\ref{h1}) and (\ref{h2}) we have

\be \langle u_{x}\Theta\rangle-i\mu_{1}\langle
uu_{x}\Theta\rangle-i\mu_{2}\langle uu_{y}\Theta\rangle
=-\lambda\frac{\partial}{\partial\mu_{1}}\langle
u\Theta\rangle=-i\lambda Z _{\mu_{1}\mu_{1}} \ee

from which we find the following expression for
$-i\alpha\mu_{1}\langle uu_{x}\Theta\rangle-i\alpha\mu_{2}\langle
uu_{y}\Theta\rangle$

\be\label{s1}
 -i\alpha\mu_{1}\langle
uu_{x}\Theta\rangle-i\alpha\mu_{2}\langle uu_{y}\Theta\rangle
=-i\alpha\lambda Z_{\mu_{1}\mu_{1}}-\alpha\langle
u_{x}\Theta\rangle. \ee

Similarly for $-i\alpha\mu_{1}\langle
vv_{x}\Theta\rangle-i\alpha\mu_{2}\langle vv_{y}\Theta\rangle$ we
find

\be\label{s2}
 -i\alpha\mu_{1}\langle
vv_{x}\Theta\rangle-i\alpha\mu_{2}\langle vv_{y}\Theta\rangle
=-i\alpha\lambda Z_{\mu_{2}\mu_{2}}-\alpha\langle
v_{y}\Theta\rangle \ee

Now from eqs. (\ref{s1}) and (\ref{s2}) and by using the Novikov's
theorem the eq.(\ref{evolut}) gives

\bea Z_{t}&=&i\gamma(t)\lambda Z+i\lambda\frac{\alpha}{2}
Z_{\mu_{1}\mu_{1}}+ i\lambda\frac{\alpha}{2} Z_{\mu_{2}\mu_{2}}
-i\alpha\lambda Z_{\mu_{1}\mu_{1}}-i\alpha\lambda Z_{\mu_{2}\mu_{2}}\nonumber\\
&-&\alpha\langle u_{x}\Theta\rangle-\alpha\langle
v_{y}\Theta\rangle-\lambda^{2}k(0,0)Z+\mu_{1}^{2}k''(0,0)Z+\mu_{2}k''(0,0)Z.
\eea

The term $\langle u_{x}\Theta\rangle$  can be evaluated as follows

\bea \langle
u_{x}\Theta\rangle&=&\frac{i}{\mu_{1}}\langle\Theta\rangle_{x}+\frac{i}{\mu_{1}}\langle
(i\lambda u+i\mu_{2}v_{x})\Theta\rangle\nonumber\\
&=&-i\frac{\lambda}{\mu_{1}}Z_{\mu_{1}}-\frac{\mu_{2}}{\mu_{1}}\langle
v_{x}\Theta\rangle, \eea

also for $\langle v_{y}\Theta\rangle$ we have

 \be \langle
v_{y}\Theta\rangle=-i\frac{\lambda}{\mu_{2}}Z_{\mu_{2}}-\frac{\mu_{1}}{\mu_{2}}\langle
u_{y}\Theta\rangle. \ee

In the appendix it is proved that $\langle
u_{y}\Theta\rangle=\langle v_{x}\Theta\rangle=0$, so the equation
governing the time evolution of $Z$ will become

\bea\label{evolut2} Z_{t}&=&i\gamma(t)\lambda
Z-i\lambda\frac{\alpha}{2} Z_{\mu_{1}\mu_{1}}-
i\lambda\frac{\alpha}{2} Z_{\mu_{2}\mu_{2}}
\nonumber\\
&+&i\alpha\frac{\lambda}{\mu_{1}}Z_{\mu_{1}}-i\alpha\frac{\lambda}{\mu_{2}}Z_{\mu_{2}}\nonumber\\
&-&\lambda^{2}k(0,0)Z+\mu_{1}^{2}k''(0,0)Z+\mu_{2}k''(0,0)Z. \eea

In what follows we solve this partial differential equation by
using flat initial condition, $h(x,y,0)=u(x,y,0)=v(x,y,0)=0$,which
means

\be P(\tilde{h},u,v,0)=\delta({\tilde h})\delta(u)\delta(v), \ee

\noindent so that the initial condition for the generating function  is
$Z(\lambda,\mu_1,\mu_2,0)=1$.\\
The solution of the
eq.(\ref{evolut2}) can be factorized as

\be\label{factor}
Z(\lambda,\mu_{1},\mu_{2},t)=F_{1}(\lambda,\mu_{1},t)F_{2}(\lambda,\mu_{2},t)
\exp{(-\lambda^{2}k(0,0)t)},
\ee

\noindent which by inserting eq.(\ref{factor}) in eq.(\ref{evolut2})
the following equation is obtained

\bea\label{evolut3}
{F_{1}}_{t}F_{2}+F_{1}{F_{2}}_{t}&=&i\gamma(t)\lambda
F_{1}F_{2}-i\lambda\frac{\alpha}{2} F_{2}{F_{1}}_{\mu_{1}\mu_{1}}-
i\lambda\frac{\alpha}{2} F_{1}{F_{2}}_{\mu_{2}\mu_{2}}
\nonumber\\
&+&i\alpha\frac{\lambda}{\mu_{1}}F_{2}{F_{1}}_{\mu_{1}}-
i\alpha\frac{\lambda}{\mu_{2}}F_{1}{F_{2}}_{\mu_{2}}\nonumber\\
&-&\lambda^{2}k(0,0)F_{1}F_{2}+\mu_{1}^{2}k''(0,0)F_{1}F_{2}+\mu_{2}k''(0,0)F_{1}F_{2}.
\eea

On the other hand, $\gamma(t)=\bar{h}_{t}=\frac{\alpha}{2}\langle
u^{2}+v^{2}\rangle$ and $\frac{\alpha}{2}\langle
u^{2}\rangle=\frac{\alpha}{2}\langle v^{2}\rangle=-\alpha
k''(0,0)t$, so eq.(\ref{evolut3}) can be separated in terms of
$\mu_{1}$ and $\mu_{2}$ and it can be seen easily that
$F_{1}(\lambda,\mu_{1},t)$ and $F_{2}(\lambda,\mu_{2},t)$ are satisfied by
the following 1+1-dimensional equation

\be\label{f}
 F_{t}(\lambda,\mu,t)=-i\lambda\frac{\alpha}{2} F_{\mu\mu}+
i\alpha\frac{\lambda}{\mu}F_{\mu}+[\mu^{2}k''(0,0)-i\alpha\lambda
k''(0,0)t]F,
 \ee

\noindent which can be solved exactly with the initial condition
$F(\lambda,\mu,0)=1$\cite{14}. The solution of eq.(\ref{f}) is

\begin{eqnarray}
&&F(\mu, \lambda , {t})= ( 1 -
{\tanh}^{2}(\sqrt{2i{{k}_{{xx}}}(0,0)\alpha \lambda}{t})) \cr \nonumber \\
&&\exp[-\frac{5}{8}\ln(1-\tanh^{4}(\sqrt{2i{{k}_{{xx}}}(0,0)\alpha
\lambda}{t}))
 \cr \nonumber \\
&+&\frac{5}{4}\tanh^{-1}(\tanh^{2}(\sqrt{2ik_{xx}(0,0)\alpha\lambda}t))-\lambda^{2}k(0,0)
t \cr \nonumber\\
&-&\frac{1}{16}\ln^{2}(\frac{1-\tanh(\sqrt{2ik_{xx}(0,0)\alpha\lambda}t)}
{1+\tanh(\sqrt{2ik_{xx}(0,0)\alpha\lambda}t)}) \cr \nonumber \\
&-&\frac{1}{2}i\mu^{2}\sqrt{\frac{2ik_{xx}(0,0)}{\alpha\lambda}}{\tanh}(\sqrt{2i{{k}_{{xx}}}(0,0)\alpha
\lambda}{t})].
\end{eqnarray}

So, the solution of eq.(\ref{evolut3}) is
\bea
&&Z(\lambda,\mu_{1},\mu_{2},t)=F(\lambda,\mu_{1},t)F(\lambda,\mu_{2},t)\exp{(-\lambda^{2}k(0,0)t)}\nonumber\\
&&= ( 1 - {\tanh}^{2}(\sqrt{2i{{k}_{{xx}}}(0,0)\alpha
\lambda}{t}))^{2}
\exp[-\frac{5}{4}\ln(1-\tanh^{4}(\sqrt{2i{{k}_{{xx}}}(0,0)\alpha
\lambda}{t}))
 \cr \nonumber \\
&&+\frac{5}{2}\tanh^{-1}(\tanh^{2}(\sqrt{2ik_{xx}(0,0)\alpha\lambda}t))-\lambda^{2}k(0,0)
t \cr \nonumber\\
&&-\frac{1}{8}\ln^{2}(\frac{1-\tanh(\sqrt{2ik_{xx}(0,0)\alpha\lambda}t)}
{1+\tanh(\sqrt{2ik_{xx}(0,0)\alpha\lambda}t)}) \cr \nonumber \\
&&-\frac{1}{2}i(\mu_{1}^{2}+\mu_{2}^{2})\sqrt{\frac{2ik_{xx}(0,0)}{\alpha\lambda}}{\tanh}(\sqrt{2i{{k}_{{xx}}}(0,0)\alpha
\lambda}{t})]. \eea

Using the generating function, the probability distribution
function (PDF) of height fluctuation can be derived by inverse
Fourier transformation as
 \bea
&&P(\tilde{h},u,v,t)=\nonumber\\
&&\int\frac{d \lambda}{2\pi}\frac{d \mu_{1}}{2\pi} \frac{d
\mu_{2}}{2\pi}
\exp{\left(i\lambda\tilde{h}+i\mu_{1}u+i\mu_{2}v
\right)}Z(\lambda,\mu_{1},\mu_{2},t).
 \eea
By expanding the solution of the generating function in powers of
$\lambda$, all the $\langle (h-{\bar h})^n \rangle $ moments can be
derived.
 For instance, the first five moments before the sharp
valley formation are

\bea \langle \tilde{h}^{2}
\rangle&=&\left(\frac{k^{2}(0,0)}{\alpha k''(0,0)}\right)^{2/3}
\left[-\frac{2}{3}(\frac{t}{t^*})^4 +2\frac{t}{t^*}\right] \cr
\nonumber \\
\langle \tilde{h}^{3} \rangle
&=&-\frac{48}{45}\left(\frac{k^{2}(0,0)}{\alpha k''(0,0)}\right)(
\frac{t}{t^*})^{6} \cr \nonumber \\
\langle\tilde{h}^{4}\rangle&=&\left(\frac{k^{2}(0,0)}{\alpha
k''(0,0)}\right)^{4/3}\left[-\frac{44}{35}(\frac{t}{t^*})^8
-8(\frac{t}{t^*})^{5}+12(\frac{t}{t^*})^{2}\right]\cr \nonumber \\
\langle\tilde{h}^{5}\rangle&=&-\left(\frac{k^{2}(0,0)}{\alpha
k''(0,0)}\right)^{5/3}\left[\frac{1216}{945}(\frac{t}{t^*})^{10}
+\frac{64}{3}(\frac{t}{t^*})^{7}\right]
\eea

\noindent where $t_* = ( \frac{ k(0,0) } { \alpha^2 k''^2 (0,0) }
) ^{1/3} $. An important information content of the derived exact
form is that they determine the time scale of the singularity
formations. One should first check the realisability condition,
i.e. $P(h-{\bar h},t) \geq 0$. In fact the moment relations
above indicate that different even order moments will get {\it
negative} in some distinct characteristic time scales. A closer
inspection into the
 even moment relations reveals that the higher the moments are,
the smaller their characteristic time scales become. So the rate
of decreasing tends to $ t_c = \frac{1}{4}t_*$ for very large
even moments asymptotically, where $t_{*}=(\frac{k(0,0)}{\alpha^2
k''^{2}(0,0)})^{1/3} = ( \frac{ \pi }{8 \alpha ^2 D_0})^{1/3}
\sigma ^2  $. Therefore, it is concluded that at this time the
right tail of the probability distribution function (PDF) starts
to become negative, which is the reminiscent of singularity
creation. Physically, the connection between the negativity of the
right tail and the singularity formation time scale is related to
a multi-valued solution of KPZ (Burgers) equation for $t \geq
t_c$. It should be notified that eq.(16) preserves the
normalisability, i.e. $Z(0,0,0,t) = 1$, equivalent to $ \int_{-
\infty} ^ {+\infty}  P(h- \bar h,u,v;t) d(h-\bar h) du dv =1$ for
every time $t$. So the PDF of $h - \bar h$ and its derivatives are
always normalizable to unity. In the $\nu = 0$ limit, for time
scales greater than  $t_c$ , the height field will become
multi-valued in the valleys (the shock region), which is related
to the $P(h-\bar h)$ left tail . The multiplicity of the height
field will increase the probability measure in the PDF. Therefore,
to compensate the exceeded measure related to the multi-valued
solutions the right tail of the PDF tails should become negative.
The singularities in the limit $\nu \rightarrow 0$  can be
constructed from multi-valued solutions of the KPZ equation with $
\nu=0$ by Maxwell cutting rule \cite{13}, which makes the
discontinuity in the derivative of the height filed.  When $t
>t_c$, the contribution of the relaxation
term in the limit of vanishing diffusion coefficient should also be considered in order to
find a realisable probability density function of height field. In
other words, the disregardance of the PDF equation's diffusion term is
only valid up to the time scales in which the singularities have not been developed
yet.

\begin{figure}
\epsfxsize=8truecm\epsfbox{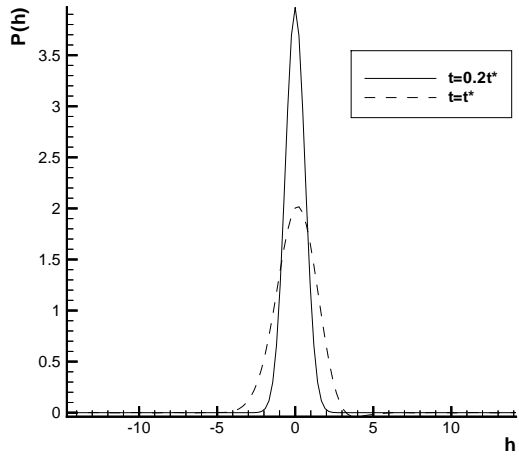}
\epsfxsize=8truecm\epsfbox{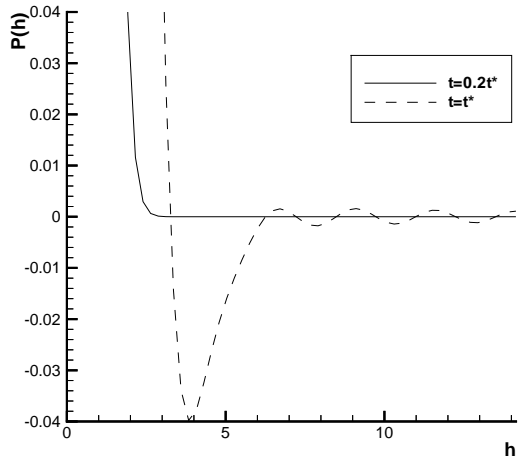} \narrowtext \caption{ In the
LHS graph the time evolution of PDF of $h-\bar h$ before
singularity formation at $\frac{2}{10} t_*$ and $ t_*$ is
numerically obtained. RHS graph shows the right tails of the PDF
of $h - \bar h$ for $\frac{2}{10} t_*$ and $ t_*$ corresponding to
before and after singularities formation which are numerically
calculated.}
\end{figure}

Taking into account $\alpha>0$ and $ k''(0,0)<0$, the odd order
moments will be  positive in time scales $t<t_c$ and the
probability density function $P(h-{\bar h},t)$ is positively
skewed. Therefore, the probability distribution functions of height
field will have a non zero skewness through its time evolution
at least before the singularity formations time scale.The inverse
Fourier transform of the generating function has been performed
numerically to obtain the PDF $P(h- \bar h)$ form. To demonstrate
the time scale of the singularity formation the PDF time
evolution has been sketched numerically, Fig(1). As the system
evolves in time, the formation of the first singularities leads
to the right tail negativity in the PDF.

To summarise, we analyse the fully nonlinear KPZ equation in 2$+$1
dimensions forced with a Gaussian noise which is white in time and
short range correlated in space. In the
 non-stationary regime when
the sharp valley structures are not yet developed we find an exact
form for the generating function of the joint fluctuations of
height and height gradient. We determine the time scale of the
sharp valley formation and the exact functional form of the time
dependence in the height difference moments at any given order. In
this paper, we have used the initial condition $ h(x,0)=0$ and $
u(x,0)=0$. This method enables us to determine the finite time
scale of singularity formation in terms of statistical properties
of the initial conditions. This calculation will be dealt with
elsewhere.  At this level we would remark on the stationary state
of the KPZ equation in the strong coupling limit. Using the
numerical results recently, it is shown that convergence to the
statistical steady state is reached after a few turnover times [16]. 
Therefore, in the stationary state the singularities will
be fully developed and, therefore, at large $t$ or $ t \rightarrow
\infty$ we should take into account the relaxation contribution in
the limit $ \nu \rightarrow 0$. Contribution of the relaxation
term in the rhs of the eq.(6) is,

\be \lim_{\nu \rightarrow 0} \{ -i\nu \mu_1 \langle\ \nabla^2 u
\Theta  \rangle -i\nu \mu_2 \langle\ \nabla^2 v  \Theta  \rangle
\}. \ee

In \cite{14} we have calculated the finite contribution of this
term in 1+1 dimensions and showed that it plays an important role
in statistical theory of KPZ equation in the stationary state. It
is also shown that all of the amplitudes of the moments $<( h -
\bar h)^n> $ can be expressed in term of the relaxation term in
the limit $ \nu \rightarrow 0$. In 2+1 dimension calculation of
finite contribution of the relaxation term is more complex (
because of complex structure of the singularities) and it is left
for a future work. We believe that the analysis followed in that
paper is also suitable for the zero temperature limit in the
problem of directed polymer in the random potential with short
range correlations [10]. In the same direction the present method
is applicable to the {\it strong coupling} regime of KPZ equation
in higher dimensions ($d > 2$) which is definitely an important
step.
\\
\\
{\bf Acknowledgement}\\
We thank J. Davoudi, F. Arash,  M. Khorrami, A. Reza-Khani and A. Tabei
 for useful discussions. F.S and A.A.M. were partially supported
 by the Institute for Studies in Physics and Mathematics (IPM).

\section{APPENDIX }
In this appendix we present another way of deriving the height
moments by introducing the second spatial derivatives of height
field. Defining
$\Theta(\lambda,\mu_{1},\mu_{2},\eta_{1},\eta_{2},\eta_{3},x,y,t)$
as

\bea
\Theta(\lambda,\mu_{1},\mu_{2},\eta_{1},\eta_{2},\eta_{3},x,y,t) =
&&\exp{
\{-i\lambda(h(x,y,t)-\bar{h}(t))-i\mu_{1}u(x,y,t)-i\mu_{2}v(x,y,t)}
\nonumber\\
&&{-i\eta_{1}w(x,y,t)-i\eta_{2}s(x,y,t)-i\eta_{3}q(x,y,t) \}} ,
 \eea

\noindent the generating function is defined as

\be Z(\lambda,\mu_{1},\mu_{2},\eta_{1},\eta_{2},\eta_{3},x,y,t)=
\langle \Theta \rangle, \ee where
$u(x,y,t)=h_{x}(x,y,t)$,$v(x,y,t)=h_{y}(x,y,t)$,$w(x,y,t)=h_{xx}(x,y,t)$,
$s(x,y,t)=h_{xy}(x,y,t)$ and $q(x,y,t)=h_{yy}(x,y,t)$.
\par
The evolution of
$h(x,y,t),u(x,y,t),v(x,y,t),w(x,y,t),s(x,y,t),q(x,y,t)$ is given
by the following equations

\bea
h_{t}&=&\frac{\alpha}{2}(u^2+v^2)+f(x,y,t)\\
u_{t}&=&\alpha(uw+vs)+f_{x}(x,y,t)\\
v_{t}&=&\alpha(us+vq)+f_{y}(x,y,t)\\
w_{t}&=&\alpha(w^2+s^2+uw_{x}+
vs_{x})+f_{xx}(x,y,t)\\
s_{t}&=&\alpha(ws+qs+vs_{y}
uw_{y})+f_{xy}(x,y,t)\\
q_{t}&=&\alpha(s^2+q^2+us_{y}+ vq_{y})+f_{yy}.
\eea
\noindent It follows from the above equations that the generating
function $Z$ is the solution of the following equation

\bea Z_{t}&=&i\gamma(t)\lambda Z-i\lambda\frac{\alpha}{2}
Z_{\mu_{1}\mu_{1}}- i\lambda\frac{\alpha}{2} Z_{\mu_{2}\mu_{2}}
+i\alpha Z_{\mu_{1}x}+i\alpha Z_{\mu_{2}y}\nonumber\\
&-&i\alpha Z_{\eta_{1}}-i\alpha Z_{\eta_{3}}
+i\alpha\eta_{1} Z_{\eta_{1}\eta_{1}}+i\alpha\eta_{1} Z_{\eta_{2}\eta_{2}}\nonumber\\
&+&i\alpha\eta_{2} Z_{\eta_{1}\eta_{2}}+i\alpha\eta_{2}
Z_{\eta_{2}\eta_{3}}
+i\alpha\eta_{3} Z_{\eta_{2}\eta_{2}}+i\alpha\eta_{3} Z_{\eta_{3}\eta_{3}}\nonumber\\
&-&i\lambda\langle f(x,y,t)\Theta \rangle -i\mu_{1}\langle
f_{x}(x,y,t)\Theta \rangle
-i\mu_{2}\langle f_{y}(x,y,t)\Theta \rangle\nonumber\\
&-&i\eta_{1}\langle f_{xx}(x,y,t)\Theta \rangle -i\eta_{2}\langle
f_{xy}(x,y,t)\Theta \rangle -i\eta_{3}\langle f_{yy}(x,y,t)\Theta
\rangle, \eea

\noindent in which $\gamma(t)$ is defined as
$\gamma(t)=\bar{h}_{t}$ and the following identities have been used

\bea \eta_{1}\langle w_{x}\Theta \rangle+\eta_{2}\langle
s_{x}\Theta \rangle +\eta_{3}\langle q_{x}\Theta
\rangle&=&iZ_{x}-i\lambda Z_{\mu_{1}}
-i\mu_{1} Z_{\eta_{1}}-i\mu_{2} Z_{\eta_{2}}\\
\eta_{1}\langle w_{y}\Theta \rangle+\eta_{2}\langle s_{y}\Theta
\rangle +\eta_{3}\langle q_{y}\Theta \rangle&=&iZ_{y}-i\lambda
Z_{\mu_{2}} -i\mu_{1} Z_{\eta_{2}}-i\mu_{2} Z_{\eta_{3}}. \eea

Now, using Novikov's theorem we find

\bea \langle f(x,y,t)\Theta \rangle&=&-i\lambda
k(0,0)Z-i\eta_{1}k_{xx}(0,0)Z
-i\eta_{3}k_{xx}(0,0)Z\\
\langle f_{x}(x,y,t)\Theta \rangle&=&i\mu_{1} k_{xx}(0,0)Z\\
\langle f_{y}(x,y,t)\Theta \rangle&=&i\mu_{2} k_{xx}(0,0)Z\\
\langle f_{xx}(x,y,t)\Theta \rangle&=&-i\lambda
k_{xx}(0,0)Z-i\eta_{1}k_{xxxx}(0,0)Z
-i\eta_{3}k_{xxxx}(0,0)Z\\
\langle f_{xy}(x,y,t)\Theta \rangle&=&-i\eta_{2} k_{xxxx}(0,0)Z\\
\langle f_{yy}(x,y,t)\Theta \rangle&=&-i\lambda
k_{xx}(0,0)Z-i\eta_{1}k_{xxxx}(0,0)Z -i\eta_{3}k_{xxxx}(0,0)Z,
\eea

\noindent where
$k(x-x',y-y')=2D_{0}D(x-x',y-y'),k(0,0)=\frac{2D_{0}}{\pi
\sigma^2} ,k_{xx}(0,0)=k_{yy}(0,0)=-\frac{4D_{0}}{\pi\sigma^4}$
and $k_{x}(0,0)=k_{y}(0,0)=0$. So we have

\bea Z_{t}&=&i\gamma(t)\lambda Z-i\lambda\frac{\alpha}{2}
Z_{\mu_{1}\mu_{1}}- i\lambda\frac{\alpha}{2} Z_{\mu_{2}\mu_{2}}
+i\alpha Z_{\mu_{1}x}+i\alpha Z_{\mu_{2}y}\nonumber\\
&-&i\alpha Z_{\eta_{1}}-i\alpha Z_{\eta_{3}}
+i\alpha\eta_{1} Z_{\eta_{1}\eta_{1}}+i\alpha\eta_{1} Z_{\eta_{2}\eta_{2}}\nonumber\\
&+&i\alpha\eta_{2} Z_{\eta_{1}\eta_{2}}+i\alpha\eta_{2}
Z_{\eta_{2}\eta_{3}}
+i\alpha\eta_{3} Z_{\eta_{2}\eta_{2}}+i\alpha\eta_{3} Z_{\eta_{3}\eta_{3}}\nonumber\\
&-&\lambda^2 k(0,0)Z
+(\mu_{1}^2+\mu_{2}^2-2\lambda\eta_{1}-2\lambda\eta_{3})
k_{xx}(0,0)Z\nonumber\\
&-&(\eta_{1}^2+\eta_{2}^2+\eta_{3}^2+2\eta_{1}\eta_{3})k_{xxxx}(0,0)Z.
\eea

Assuming statistical homogeneity $(Z_{x}=0,Z_{y}=0)$ and defining
$P(\tilde{h},u,v,w,s,q,t)$ as the joint probability density
function of $\tilde{h},u,v,w,s$ and $q$, one can construct the PDF
as the Fourier transform
 of the generating function $Z$ with respect to
 $\lambda,\mu_{1},\mu_{2},\eta_{1},\eta_{2},\eta_{3}$

\bea
&&P(\tilde{h},u,v,w,s,q,t)=\int\frac{d \lambda}{2\pi}\frac{d
\mu_{1}}{2\pi} \frac{d \mu_{2}}{2\pi}\frac{d
\eta_{1}}{2\pi}\frac{d \eta_{2}}{2\pi}
\frac{d \eta_{3}}{2\pi}\{\nonumber\\
&& \exp{\left(i\lambda\tilde{h}+i\mu_{1}u+i\mu_{2}v
+i\eta_{1}w+i\eta_{2}s+i\eta_{3}q\right)}Z(\lambda,\mu_{1},\mu_{2},\eta_{1}
,\eta_{2},\eta_{3},t)\}.\hspace{2cm} \eea

From the eqs. (43) and (44) the equation governing the
evolution of $P(\tilde{h},u,v,w,s,q,t)$ can be derived, which is

\bea
P_{t}&=&\gamma(t)P_{\tilde{h}}+\frac{\alpha}{2}(u^2+v^2)P_{\tilde{h}}-4\alpha
wP
-4\alpha qP\nonumber\\
&-&\alpha w^2P_{w}-\alpha q^2P_{q}-\alpha s^2P_{w}-\alpha s^2P_{q}
-\alpha wsP_{s}-\alpha qsP_{s}\nonumber\\
&+&k(0,0)P_{\tilde{h}\tilde{h}}+2k_{xx}(0,0)P_{\tilde{h}w}+2k_{xx}(0,0)P_{\tilde{h}q}
-k_{xx}(0,0)P_{uu}-k_{xx}(0,0)P_{vv}\nonumber\\
&+&k_{xxxx}(0,0)P_{ww}+k_{xxxx}(0,0)P_{ss}+
k_{xxxx}(0,0)P_{qq}+2k_{xxxx}(0,0)P_{qw}.
 \eea

From the eq.(45), it is easy to see that the moments $\langle
\tilde{h}^{n_{0}} u^{n_{1}} v^{n_{2}} w^{n_{3}} s^{n_{4}}
q^{n_{5}} \rangle$ satisfy  the following equation
\newpage
\bea
&&\frac{d}{dt} \langle \tilde{h}^{n_{0}} u^{n_{1}} v^{n_{2}}
w^{n_{3}} s^{n_{4}} q^{n_{5}} \rangle= -n_{0}\gamma(t)\langle
\tilde{h}^{n_{0}-1} u^{n_{1}} v^{n_{2}} w^{n_{3}} s^{n_{4}}
q^{n_{5}}\rangle -\frac{\alpha n_{0}}{2}\langle
\tilde{h}^{n_{0}-1} u^{n_{1}+2} v^{n_{2}}
w^{n_{3}} s^{n_{4}} q^{n_{5}}\rangle \nonumber\\
&&-\frac{\alpha n_{0}}{2}\langle \tilde{h}^{n_{0}-1} u^{n_{1}}
v^{n_{2}+2} w^{n_{3}} s^{n_{4}} q^{n_{5}}\rangle -4\alpha \langle
\tilde{h}^{n_{0}} u^{n_{1}} v^{n_{2}} w^{n_{3}+1} s^{n_{4}}
q^{n_{5}}\rangle \nonumber\\
&&-4\alpha \langle \tilde{h}^{n_{0}} u^{n_{1}} v^{n_{2}}
w^{n_{3}} s^{n_{4}} q^{n_{5}+1}\rangle +\alpha(n_{3}+2)\langle
\tilde{h}^{n_{0}} u^{n_{1}} v^{n_{2}} w^{n_{3}+1} s^{n_{4}}
q^{n_{5}}\rangle \nonumber\\
&&+\alpha n_{3}\langle
\tilde{h}^{n_{0}} u^{n_{1}} v^{n_{2}} w^{n_{3}-1} s^{n_{4}+2}
q^{n_{5}}\rangle +\alpha(n_{4}+1)\langle \tilde{h}^{n_{0}}
u^{n_{1}} v^{n_{2}}
w^{n_{3}+1} s^{n_{4}} q^{n_{5}}\rangle  \nonumber\\
&&+\alpha(n_{4}+1)\langle \tilde{h}^{n_{0}} u^{n_{1}} v^{n_{2}}
w^{n_{3}} s^{n_{4}} q^{n_{5}+1}\rangle +\alpha(n_{5}+2)\langle
\tilde{h}^{n_{0}} u^{n_{1}} v^{n_{2}} w^{n_{3}} s^{n_{4}}
q^{n_{5}+1}\rangle\nonumber\\
&&+n_{5}(n_{5}-1)k_{xxxx}(0,0)\langle \tilde{h}^{n_{0}} u^{n_{1}}
v^{n_{2}} w^{n_{3}} s^{n_{4}}
q^{n_{5}-2}\rangle+2n_{3}n_{5}k_{xxxx}(0,0)\langle
\tilde{h}^{n_{0}} u^{n_{1}} v^{n_{2}} w^{n_{3}-1} s^{n_{4}}
q^{n_{5}-1}\rangle  \nonumber\\
&&+2n_{0}n_{3}k_{xx}(0,0)\langle \tilde{h}^{n_{0}-1} u^{n_{1}}
v^{n_{2}} w^{n_{3}-1} s^{n_{4}} q^{n_{5}}\rangle
+2n_{0}n_{5}k_{xx}(0,0)\langle \tilde{h}^{n_{0}-1} u^{n_{1}}
v^{n_{2}} w^{n_{3}} s^{n_{4}} q^{n_{5}-1}\rangle \nonumber\\
&&-n_{1}(n_{1}-1)k_{xx}(0,0)\langle \tilde{h}^{n_{0}} u^{n_{1}-2}
v^{n_{2}} w^{n_{3}} s^{n_{4}} q^{n_{5}}\rangle
-n_{2}(n_{2}-1)k_{xx}(0,0)\langle \tilde{h}^{n_{0}} u^{n_{1}}
v^{n_{2}-2} w^{n_{3}} s^{n_{4}} q^{n_{5}}\rangle \nonumber\\
&&+n_{3}(n_{3}-1)k_{xxxx}(0,0)\langle \tilde{h}^{n_{0}} u^{n_{1}}
v^{n_{2}} w^{n_{3}-2} s^{n_{4}} q^{n_{5}}\rangle
+n_{4}(n_{4}-1)k_{xxxx}(0,0)\langle \tilde{h}^{n_{0}} u^{n_{1}}
v^{n_{2}} w^{n_{3}} s^{n_{4}-2} q^{n_{5}}\rangle\nonumber\\
&& +\alpha n_{5}\langle \tilde{h}^{n_{0}} u^{n_{1}} v^{n_{2}}
w^{n_{3}} s^{n_{4}+2} q^{n_{5}-1}\rangle
+n_{0}(n_{0}-1)k(0,0)\langle \tilde{h}^{n_{0}-2} u^{n_{1}}
v^{n_{2}} w^{n_{3}} s^{n_{4}} q^{n_{5}}\rangle. \eea

By substituting different sort of  values for
$n_{0},n_{1},n_{2},n_{3},n_{4},n_{5}$ we can find some coupled
differential equations for different moments. For example, if we
take $n_{0}=n_{1}=\cdot \cdot \cdot=n_{5}=0$ we have

\be <w>+<q>=0 \ee or \be <\nabla \cdot \bf{u}>=0 \ee

\noindent where this is the same as the statistical hemogeneity
condition. For $n_{0}=1,n_{1}=n_{2}=n_{3}=n_{4}=n_{5}=0$ we find

\be <\tilde h w>+<\tilde h q>=-<u^2>-<v^2> \ee

\noindent and for $n_{0}=0,n_{1}=1,n_{2}=n_{3}=n_{4}=n_{5}=0$ we find

\be <u w>+<u q>=0, \ee

\noindent and if we assume the statistical hemogeneity we have

\be <uw>=<uu_{x}>=\frac{1}{2}<u^2>_{x}=0. \ee

\noindent So $<uw>=0$ and then $<uq>=0$, also for
$n_{0}=n_{1}=0,n_{2}=1,n_{3}=n_{4}=n_{5}=0$ we find
$<vq>=0,<vw>=0$.

For $n_{0}=0,n_{1}=2,n_{2}=n_{3}=n_{4}=n_{5}=0$ we find

\be \frac{d}{dt}<u^2>=-\alpha <u^2w>-\alpha <u^2q>-2k_{xx}(0,0)
\ee

\noindent and also we have $<u^2w>=\frac{1}{3}<u^3>_{x}=0$ by
statistical hemogeneity and also
\newline$<u^2q>=<u^2v>_{y}-2<uvs>$,
$<u^2v>_{y}=0$, so $<u^2q>=-2<uvs>$ so that we have

\be \frac{d}{dt}<u^2>=2\alpha <uvs>-2k_{xx}(0,0). \ee

\noindent The corresponding differential equation for $<uvs>$ is

\be \frac{d}{dt}<uvs>=0. \ee

\noindent If we assume that at $t=0$ the surface is flat so all
the moments at $t=0$ are zero,so that $<uvs>=0$ and then we find

\be <u^2>=-2k_{xx}(0,0)t. \ee

\noindent Similar calculations give

\be <v^2>=-2k_{xx}(0,0)t. \ee

By focusing on the differential equation of the PDF it is deduced  that this
equation is invariant under $u\rightarrow -u$ and $v\rightarrow
-v$ which is a consequence of the inversion and reflection symmetry
of the KPZ equation. So it will result in  the vanishing of the $u$ and $v$
odd moments
\be \langle u^{2k+1}\rangle=\langle v^{2k+1}\rangle=0. \ee

\noindent Using eq.(46) the even moments of $u$ and $v$ can be
calculated. For example, for $\langle u^4 \rangle$ we have

\be \frac{d}{dt}\langle u^4 \rangle=-\alpha\langle u^{4}w \rangle
-\alpha\langle u^{4}q \rangle-12k_{xx}(0)\langle u^2 \rangle. \ee
On the other hand $\langle u^{4}w \rangle$ can be written  as

\be \langle u^{4}u_{x} \rangle=\frac{1}{5}\langle u^{5}
\rangle_{x}, \ee
which is zero by homogeneity, also for $\langle
u^{4}q \rangle$ we have

\be \langle u^{4}v_{y} \rangle=\langle u^{4}v \rangle_{y}
-4\langle u^{3}vs \rangle=-4\langle u^{3}vs \rangle \ee in which
$\langle u^{4}v \rangle_{y}$ is zero by homogeneity. The
differential equation for $\langle u^{3}vs \rangle$ is

\be \frac{d}{dt}\langle u^{3}vs \rangle=0, \ee so $\langle u^{3}vs
\rangle=0$. Therefore, it is obtained

\be \langle u^{4} \rangle=12k_{xx}^{2}(0,0)t^2, \ee
and we have  the same
for $\langle v^{4} \rangle$. Also it can be found that $\langle
u^{2}v^{2} \rangle=4k_{xx}^{2}(0,0)t^2$. By continuing the above
method all the moments $\langle u^{n}v^{m}\rangle$ can be deduced.

On the other hand, the above calculations show that  all the mixed moments
$\langle u^{n}v^{m}s \rangle$ are zero. In the following  some of
the moments of $u$ and $v$ and their combinations have been given

\bea
\langle u^{6} \rangle&=&\langle v^{6} \rangle=-120k_{xx}^{3}(0,0)t^3\\
\langle u^{2}v^{4} \rangle&=&\langle u^{4}v^{2} \rangle=-24k_{xx}^{3}(0,0)t^3\\
\langle u^{8} \rangle&=&\langle v^{8} \rangle=1680k_{xx}^{4}(0,0)t^4\\
\langle u^{6}v^{2} \rangle&=&\langle u^{2}v^{6} \rangle=240k_{xx}^{4}(0,0)t^{4}\\
\langle u^{4}v^{4} \rangle&=&144k_{xx}^{4}(0,0)t^{4} , etc.
\eea

We aim to calculate all the momentes of $\tilde{h}=h-\bar{h}$.
Putting $n_{0}=2,n_{1}=n_{2}=\cdot\cdot\cdot=0$ in the eq.(46), we
find

\be \frac{d}{dt}\langle \tilde{h}^{2} \rangle=-\alpha\langle
\tilde{h}u^{2} \rangle -\alpha\langle \tilde{h}v^{2}
\rangle-\alpha\langle \tilde{h}^{2}w \rangle -\alpha\langle
\tilde{h}^{2}q \rangle+2k(0,0). \ee We know that $\langle
\tilde{h}^{2}w \rangle=\langle \tilde{h}^{2}u \rangle_{x}-
2\langle \tilde{h}u^{2} \rangle$ and because of the homogeneity
$\langle \tilde{h}^{2}u \rangle_{x}=0$, so we have

\be \langle \tilde{h}^{2}w \rangle=-2\langle \tilde{h}u^{2}
\rangle, \ee and also

\be \langle \tilde{h}^{2}q \rangle=-2\langle \tilde{h}v^{2}
\rangle, \ee then \be \frac{d}{dt}\langle \tilde{h}^{2}
\rangle=\alpha (\langle \tilde{h}u^{2} \rangle+\langle
\tilde{h}v^{2} \rangle)+2k(0,0). \ee

As it can be seen from eq.(71) we need the moments $\langle
\tilde{h}u^{2} \rangle$ and $\langle \tilde{h}v^{2} \rangle$ to
find  $\langle \tilde{h}^{2} \rangle$ , which the related
differential equations are

\bea \frac{d}{dt}\langle \tilde{h}u^{2}
\rangle&=&-\gamma(t)\langle u^{2} \rangle -\frac{\alpha}{2}\langle
u^{4} \rangle-\frac{\alpha}{2} \langle u^{2}v^{2}
\rangle-\alpha\langle \tilde{h}u^{2}w \rangle
-\alpha\langle \tilde{h}u^{2}q \rangle \\
\frac{d}{dt}\langle \tilde{h}v^{2} \rangle&=&-\gamma(t)\langle
v^{2} \rangle -\frac{\alpha}{2}\langle v^{4}
\rangle-\frac{\alpha}{2} \langle u^{2}v^{2} \rangle-\alpha\langle
\tilde{h}v^{2}w \rangle -\alpha\langle \tilde{h}v^{2}q \rangle .
\eea

By using of the statistical homogeneity the last
two terms of the above equations can be converted
as

\bea
\langle \tilde{h}u^{2}w \rangle&=&-\frac{1}{3}\langle u^{4} \rangle\nonumber\\
\langle \tilde{h}v^{2}q \rangle&=&-\frac{1}{3}\langle v^{4} \rangle \nonumber\\
\langle \tilde{h}u^{2}q \rangle&=&-\langle u^{2}v^{2} \rangle
-2\langle \tilde{h}uvs \rangle \nonumber\\
\langle \tilde{h}v^{2}w \rangle&=&-\langle u^{2}v^{2} \rangle
-2\langle \tilde{h}uvs \rangle. \eea

\noindent The above relations result in

\bea \frac{d}{dt}(\langle \tilde{h}u^{2} \rangle+\langle
\tilde{h}v^{2} \rangle&=& -\gamma(t)(\langle u^{2}+v^{2} \rangle)
-\frac{\alpha}{6}(\langle u^{4}+v^{4} \rangle)\nonumber\\
&+&\alpha\langle u^{2}v^{2} \rangle +4\alpha\langle \tilde{h}uvs
\rangle. \eea

\noindent Obtaining the differential equation for $\langle
\tilde{h}uvs \rangle$ we get

\be \frac{d}{dt}\langle \tilde{h}uvs \rangle=0, \ee

\noindent which results in $\langle \tilde{h}uvs \rangle=0$. Also
it is easy to see that all the moments $\langle
\tilde{h}^{n}u^{m}v^{p}s \rangle$ are zero. This fantastic result
helps us to find all the $\langle \tilde{h}^{n} \rangle$ moments.
Now by substituting $\gamma(t)=\bar
{h}_{t}=\frac{\alpha}{2}(\langle u^{2}+v^{2} \rangle)=-2\alpha
k_{xx}(0,0)t$ , $\langle u^{4} \rangle$ , $\langle v^{4} \rangle$
and $\langle u^{2}v^{2} \rangle$ in eq.(75) we obtain

\be
\langle \tilde{h}u^{2} \rangle+\langle \tilde{h}v^{2} \rangle=
-\frac{8}{3}\alpha k_{xx}^{2}(0,0)t^3,
\ee
 which finally gives

\be \langle \tilde{h}^{2}
\rangle=-\frac{2}{3}\alpha^{2}k_{xx}^{2}(0,0)t^4 +2k(0,0)t. \ee

Now we begin to calculate the moment $\langle \tilde{h}^{3}
\rangle$. Inserting $n_{0}=3,n_{1}=n_{2}=\cdot\cdot\cdot=n_{5}=0$
in eq.(46) we get

\be \frac{d}{dt}\langle \tilde{h}^{3} \rangle=-3\gamma(t)\langle
\tilde{h}^{2} \rangle -\frac{3}{2}\alpha(\langle
\tilde{h}^{2}u^{2} \rangle+ \langle \tilde{h}^{2}v^{2}
\rangle)-\alpha\langle \tilde{h}^{3}w \rangle -\alpha\langle
\tilde{h}^{3}q \rangle, \ee
and again by using statisiacal homogeneity it
can be shown that

\bea
\langle \tilde{h}^{3}w \rangle&=&-3\langle \tilde{h}^{2}u^{2} \rangle \\
\langle \tilde{h}^{3}q \rangle&=&-3\langle \tilde{h}^{2}v^{2}
\rangle. \eea

\noindent So we have

\be \frac{d}{dt}\langle \tilde{h}^{3} \rangle=-3\gamma(t)\langle
\tilde{h}^{2} \rangle +\frac{3}{2}\alpha(\langle
\tilde{h}^{2}u^{2} \rangle+ \langle \tilde{h}^{2}v^{2} \rangle).
\ee To calculate $\langle \tilde{h}^{3} \rangle$ we need
$\langle \tilde{h}^{2} \rangle$ , $(\langle \tilde{h}^{2}u^{2}+
\tilde{h}^{2}v^2 \rangle )$. $\langle \tilde{h}^{2} \rangle$ has
been calculated above, so we will obtain $\langle
\tilde{h}^{2}u^{2}+\tilde{h}^{2}v^{2}  \rangle$ using the corresponding differential equation as follows,

\bea \frac{d}{dt}(\langle \tilde{h}^{2}u^{2} \rangle+ \langle
\tilde{h}^{2}v^{2} \rangle&=& -2\gamma(t)(\langle
\tilde{h}u^{2}+\tilde{h}v^{2} \rangle) -\alpha(\langle
\tilde{h}u^{4}+\tilde{h}v^{4} \rangle) -2\alpha\langle
\tilde{h}u^{2}v^{2} \rangle
-\alpha(\langle \tilde{h}^{2}u^{2}w+ \tilde{h}^{2}v^{2}w \rangle)\nonumber\\
&-&\alpha(\langle \tilde{h}^{2}u^{2}q+ \tilde{h}^{2}v^{2}q
\rangle) +2k(0,0)(\langle u^{2}+ v^{2} \rangle)
-4k_{xx}(0,0)\langle \tilde{h}^{2} \rangle. \eea

As before, we easily get

\bea \langle \tilde{h}^{2}u^{2}w \rangle &=&
-\frac{2}{3}\langle \tilde{h}u^{4}\rangle\nonumber\\
\langle \tilde{h}^{2}v^{2}q \rangle&=&
-\frac{2}{3}\langle \tilde{h}v^{4} \rangle\nonumber\\
\langle \tilde{h}^{2}u^{2}q\rangle&=&
-2\langle \tilde{h}u^{2}v^{2} \rangle-2\langle \tilde{h}^{2}uvs\rangle \nonumber\\
\langle \tilde{h}^{2}v^{2}w\rangle&=& -2\langle
\tilde{h}u^{2}v^{2} \rangle-2\langle \tilde{h}^{2}uvs\rangle .
\eea

As discussed before, it is easy to show that the moment $\langle
\tilde{h}^{2}uvs\rangle$ is zero. To prove this we write the
corresponding differential equation

\be \frac{d}{dt}\langle \tilde{h}^{2}uvs\rangle= -\alpha(\langle
\tilde{h}u^{3}vs \rangle+\langle \tilde{h}uv^{3}s\rangle ), \ee
and again by trying to write the differential eqautions for $\langle
\tilde{h}u^{3}vs \rangle$ and $\langle \tilde{h}uv^{3}s \rangle$
we obtain

\bea
\frac{d}{dt}\langle \tilde{h}u^{3}vs \rangle&=&0\nonumber\\
\frac{d}{dt}\langle \tilde{h}uv^{3}s \rangle&=&0 \eea
which
results in $\langle \tilde{h}u^{3}vs \rangle=\langle
\tilde{h}uv^{3}s \rangle=0$, therefore $\langle \tilde{h}^{2}uvs
\rangle=0$. Now $\langle \tilde{h}u^{4} \rangle$ ,
$\langle \tilde{h}v^{4} \rangle$ and $\langle \tilde{h}u^{2}v^{2}
\rangle$ should be found. The relating differential equation for $\langle
\tilde{h}u^{4}+\tilde{h}v^{4} \rangle$ is

\bea \frac{d}{dt}(\langle \tilde{h}u^{4} \rangle+ \langle
\tilde{h}v^{4} \rangle&=& -\gamma(t)(\langle u^{4}+v^{4} \rangle)
-\frac{\alpha}{2}(\langle u^{6}+v^{6} \rangle)
-\frac{\alpha}{2}(\langle u^{4}v^{2}+u^{2}v^{4} \rangle)\nonumber\\
&-&\alpha(\langle \tilde{h}u^{4}w+ \tilde{h}v^{4}w \rangle)
-\alpha(\langle \tilde{h}u^{4}q+ \tilde{h}v^{4}q \rangle)
-12k_{xx}(0,0)(\langle \tilde{h}u^{2}+\tilde{h}v^{2} \rangle).
\eea

As before the following identities are held

\bea
\langle \tilde{h}u^{4}w \rangle &=& -\frac{1}{5}\langle u^{6} \rangle \nonumber\\
\langle \tilde{h}u^{4}q \rangle &=& -\langle u^{4}v^{2} \rangle \nonumber\\
\langle \tilde{h}v^{4}q \rangle &=& -\frac{1}{5}\langle v^{6} \rangle \nonumber\\
\langle \tilde{h}v^{4}w \rangle &=& -\langle u^{2}v^{4} \rangle,
\eea

\noindent so we have

\bea \frac{d}{dt}(\langle \tilde{h}u^{4} \rangle+ \langle
\tilde{h}v^{4} \rangle&=& -\gamma(t)(\langle u^{4}+v^{4} \rangle)
-\frac{3\alpha}{10}(\langle u^{6}+v^{6} \rangle)
+\frac{\alpha}{2}(\langle u^{4}v^{2}+u^{2}v^{4} \rangle)\nonumber\\
&-&12k_{xx}(0,0)(\langle \tilde{h}u^{2}+\tilde{h}v^{2} \rangle).
\eea Substituting the expressions for $\gamma(t)$ ,$\langle
u^{4}\rangle$ , $\langle v^{4} \rangle$ , $\langle u^{6} \rangle$
, $\langle v^{6} \rangle$ , $\langle u^{4}v^{2} \rangle$, $\langle
u^{2}v^{4} \rangle$ and $(\langle \tilde{h}u^{2}+\tilde{h}v^{2}
\rangle)$

we find

\be \langle \tilde{h}u^{4}+\tilde{h}v^{4} \rangle=32\alpha
k_{xx}(0,0)^{3}t^{4}. \ee

The corresponding differential equation for $\langle
\tilde{h}u^{2}v^{2} \rangle$ is

\bea \frac{d}{dt}(\langle \tilde{h}u^{2}v^{2} \rangle&=&
-\gamma(t)(\langle u^{2}v^{2} \rangle)
-\frac{\alpha}{2}(\langle u^{4}v^{2}+u^{2}v^{4} \rangle)\nonumber\\
&-&\alpha(\langle \tilde{h}u^{2}v^{2}w+\tilde{h}u^{2}v^{2}q
\rangle) -2k_{xx}(0,0)(\langle \tilde{h}u^{2}+\tilde{h}v^{2}
\rangle). \eea

As before the following identities are held

\bea \langle \tilde{h}u^{2}v^{2}w \rangle &=&
-\frac{1}{3}\langle u^{4}v^{2} \rangle\nonumber\\
\langle \tilde{h}u^{2}v^{2}q \rangle &=& -\frac{1}{3}\langle
u^{2}v^{4} \rangle, \eea

\noindent then by inserting the known moments we find

\be \langle \tilde{h}u^{2}v^{2} \rangle = \frac{16}{3}\alpha
k_{xx}^{3}(0,0)t^4. \ee

The above results get

\be \langle \tilde{h}^{2}u^{2} \rangle+\langle \tilde{h}^{2}v^{2}
\rangle=
-\frac{8}{3}\alpha^{2}k_{xx}^{3}(0,0)t^{5}-8k(0,0)k_{xx}(0,0)t^{2}
, \ee

\noindent which finally results in

\be \langle \tilde{h}^{3}
\rangle=-\frac{48}{45}\alpha^{3}k_{xx}^{3}(0,0)t^{6}. \ee

By continuing the above proccedure all the $\langle\tilde{h}^{n}\rangle$ moments can be derived. Some of these moments are listed
bellow

\bea
\langle\tilde{h}^{4}\rangle&=&-\frac{44}{35}\alpha^{4}k_{xx}^{4}(0,0)t^{8}
-8\alpha^{2}k(0,0)k_{xx}^{2}(0,0)t^{5}+12k^{2}(0,0)t^{2}\\
\langle\tilde{h}^{5}\rangle&=&-\frac{1216}{945}\alpha^{5}k_{xx}^{5}(0,0)t^{10}
-\frac{64}{3}\alpha^{3}k(0,0)k_{xx}^{3}(0,0)t^{7},
\eea

\noindent which are the same as the  results that we derived  by
expanding the generating function.
 Results of this appendix show that any moment containing the
 first power of $s$ vanishes. Indeed, one can prove the following
 identity

 \be
 \langle s e^{-i(\lambda {\tilde h}+\mu_{1}u+\mu_{2}v)}\rangle=0.
 \ee

By expanding the exponential in the above expression one finds

\be \langle s e^{-i(\lambda {\tilde
h}+\mu_{1}u+\mu_{2}v)}\rangle=\sum_{n,m,p}\frac{i^{(n+m+p)}\lambda^n
\mu_1^m \mu_2^p }{n !m!p!}\langle s{\tilde
h}^{n}u^{m}v^{p}\rangle. \ee

Now setting $n_{3}=n_{5}=0$ and $n_{4}=1$ in eq.(46) we get the
following equation for $\langle s{\tilde h}^{n}u^{m}v^{p}\rangle$

 \bea \frac{d}{d t} \langle {\tilde h}^{n} u^{m} v^{p}s
\rangle= &-&n\gamma(t)\langle {\tilde h}^{n-1} u^{m} v^{p}s\rangle
-\frac{\alpha n}{2}\langle {\tilde h}^{n-1} u^{m+2} v^{p} s
\rangle -\frac{\alpha n}{2}\langle {\tilde h}^{n-1} u^{m} v^{p+2}
s\rangle \cr\nonumber\\
&+& n(n-1)k_{xx}(0,0)\langle {\tilde h}^{n-2} u^{m} v^{p} s\rangle
-m(m-1)k_{xx}(0,0)\langle {\tilde h}^{n} u^{m-2} v^{p}s\rangle \cr\nonumber\\
&-&p(p-1)k_{xx}(0,0)\langle {\tilde h}^{n} u^{m} v^{p-2} s\rangle.
\eea

Starting from eqs.(54),(61) and (76) all the $\langle s{\tilde
h}^{n}u^{m}v^{p}\rangle$ moments can be evaluated and it can be
shown that, with flat initial condition, they are equal to zero.
So we conclude $\langle s\Theta\rangle=\langle
u_{y}\Theta\rangle=\langle v_{x}\Theta\rangle=0$.

\end{document}